\documentclass[draftcls, onecolumn, 11pt]{IEEEtran}

\usepackage{hyperref,bm,graphicx,amsmath,amssymb}
\usepackage{algorithm,algpseudocode}

\usepackage{color, soul}

\newcommand{\FCR}[1]{\textcolor{red}{#1}}

\begin{document}
\title{ On Detection-Directed Estimation Approach \\for Noisy Compressive Sensing }
\bigskip

\author{\IEEEauthorblockN{Jaewook Kang, Heung-No Lee, and *Kiseon
Kim}\\
\bigskip
\IEEEauthorblockA{School of Information and Communication,\\
 Department of Nanobio Materials and Electronics,\\
 Gwangju Institute of Science and Technology (GIST), Gwangju 500-712, South Korea\\
(Tel.: +82-62-715-2264, Fax.:+82-62-715-2274,
Email:\{jwkkang,heungno,*kskim\}@gist.ac.kr}) }

 \maketitle

\begin{abstract}
In this paper, we investigate a Bayesian sparse reconstruction
algorithm called compressive sensing via Bayesian support detection
(CS-BSD). This algorithm is quite robust against measurement noise
and achieves the performance of an minimum mean square error (MMSE)
estimator that has support knowledge beyond a certain SNR thredhold.
The key idea behind CS-BSD is that reconstruction takes a
detection-directed estimation structure consisting of two parts:
support detection and signal value estimation. Belief propagation
(BP) and a Bayesian hypothesis test perform support detection and an
MMSE estimator finds the signal values belonging to the support set.
CS-BSD converges faster than other BP-based algorithms and it can be
converted to an parallel architecture to become much faster.
Numerical results are provided to verify the superiority of CS-BSD,
compared to recent algorithms.
\end{abstract}


\section{Introduction}

Compressive sensing (CS) in the presence of noise has been
intensively investigated in many recent papers because any
real-world device is subject to at least a small amount of noise. We
refer to such problems as \emph{noisy compressive sensing} (NCS).
Let $\mathbf{x}=[x_1,...,x_N]$ denote a random vector whose elements
are sparsely non-zeros, called \emph{sparse signal}. Then, the NCS
decoder observes a measurement vector $\mathbf{z}=[z_1,...,z_M] \in
\mathbb{R}^M$, given as
\begin{eqnarray}\label{eq:eq_1}
\mathbf{z}=\mathbf{\Phi}\mathbf{x}_0 + \mathbf{n},
\end{eqnarray}
where $\mathbf{x}_0 \in \mathbb{R}^N $ is a deterministic sparse
signal; $\mathbf{\Phi} \in \mathbb{R}^{M \times N}$ is a sensing
matrix whose columns represent a possibly overcomplete basis,
\emph{i.e.}, rank($\mathbf{\Phi}) \leq M$, where $M <N$;and
$\mathbf{n} \in \mathbb{R}^M$ is an additive noise vector generated
by a certain distribution.

The NCS reconstruction problem has been discussed in terms of
conventional $l_1$-norm approaches \cite{Donoho2}-\cite{candes2}. In
\cite{Donoho2}-\cite{tropp}, the authors assume a bounded noise and
in \cite{Haupt},\cite{candes2}, an i.i.d. zero-mean Gaussian noise
is assumed, \emph{i.e.},  $\mathbf{n} \sim
\mathcal{N}(0,\sigma_n^2\mathbf{I}_M )$. In \cite{candes2}, Candes
and Tao proposed an $l_1$-norm based reconstruction algorithm for
the Gaussian setup, called the \emph{Dantzig selector} (L1-DS):
\begin{eqnarray}\label{eq:eq_2}
\widehat{\bf{x}} = \arg \mathop {\min }\limits_{\bf{x}} \left\|
{\bf{x}} \right\|_1 \,\,\,\,\,\,s.t.\,\, \mathbf{E} \left\|
{{\bf{\Phi }}^* ({\bf{\Phi x}} - {\bf{z}})} \right\|_\infty   \le
\epsilon ,
\end{eqnarray}
where $\epsilon$ is the tolerance user defined paramter and $*$
denotes matrice tranposition.  The reconstruction performance of
L1-DS is proprtional to logarithmic factor, \emph{i.e.}, $\mathbf{E}
\left\| {\widehat{\bf{x}} - {\bf{x}}_0} \right\|_2^2 \le C \cdot
\sigma _n^2 K \left( {\log N} \right)$ with a constant $C$ (see Th.1
in \cite{candes2}).

Alternatively, Bayesian approaches to NCS have received attention
\cite{SBL}-\cite{SuPrEM}. This type of approach offers powerful
mitigation of noise effects by using many existing statistical
signal processing techniques and several statistical signal-noise
models. In these approaches, the reconstruction problem is described
as the \emph{maximum a posteriori} (MAP) estimation problem as
follows:
\begin{eqnarray} \label{eq:eq_3}
\mathbf{\widehat{x}}=\arg\mathop {\max }\limits_{\mathbf{x}}
\,\,f_{\mathbf{x}}( \mathbf{x} | \mathbf{z} )
\,\,\,\,\,\,s.t.\,\,\,\mathbf{E} \left\| {{\bf{\Phi x }} - {\bf{z}}}
\right\|_2 \le \epsilon,
\end{eqnarray}
where Gaussian noise is assumed and  $f(\cdot)$ is a probability
density function.

The most well-known Bayesian approach is the \emph{sparse Bayesian
learning} (SBL) algorithm \cite{SBL}-\cite{BCS-LP}. The SBL
algorithm iteratively determines the posterior density of the signal
on basis of a three-layer hierarchical prior model, so the prior
density is a function of certain parameters. The algorithm estimates
the parameters of the prior using \emph{expectation maximization}
(EM) and applies these parameters to finding the posterior . The SBL
approach to sparse reconstruction was originally proposed in
\cite{SBL},\cite{SBL-BS}. Recently, Ji \emph{et al.} \cite{BCS} and
Babacan \emph{et al.} \cite{BCS-LP} successfully applied the SBL
approach to the NCS reconstruction problem with different prior
model.

Another class of  Bayesian approaches is sparse reconstruction using
sparse matrices \cite{CS-BP1}-\cite{SuPrEM}. The work is inspired by
the success of \emph{low-density parity-check} (LDPC) codes in
channel coding field \cite{Gallager}-\cite{Mackey}.  The use of the
sparse matrix enables simple and fast signal acquisition  that is
feasible in real-world applications. In addition, these approaches
can be made more attractive if they are applied in in conjunction
with belief propagation (BP). BP replaces the reconstruction process
by iterative message-passing processes. This replacement reduces the
reconstruction complexity to the $O(N \log N)$ order.

Baron \emph{et al.} for the first time proposed the use of sparse
matrices to the NCS setup and developed a BP-based algorithm, called
CS-BP \cite{CS-BP1},\cite{CS-BP2}. CS-BP iteratively updates the
signal posterior from the two-state Gaussian mixture prior via the
message-passing algorithm, where the messages are the probability
densities of the signal elements. In \cite{BP-SBL}, Tan \emph{et
al.} proposed another BP-based algorithm called, BP-SBL. They
applied BP to the SBL-framework in \cite{BCS} to reduce the
complexity of the EM algorithm. Most recently, Akcakaya \emph{et
al.} devised SuPrEM using an idea  similar to BP-SBL, but in a
different framework \cite{SuPrEM} which is based on Gaussian scale
mixture \cite{GSM} with a specific type of prior called the
Jeffreys' prior \cite{Jeffreys}. In addition, the authors restrict
the story of SuPrEM to a class of sensing matrices, called
low-density frames, in which the matrices have fixed column and row
weights.

\begin{figure}[!t]
\centering
\includegraphics[width=13 cm]{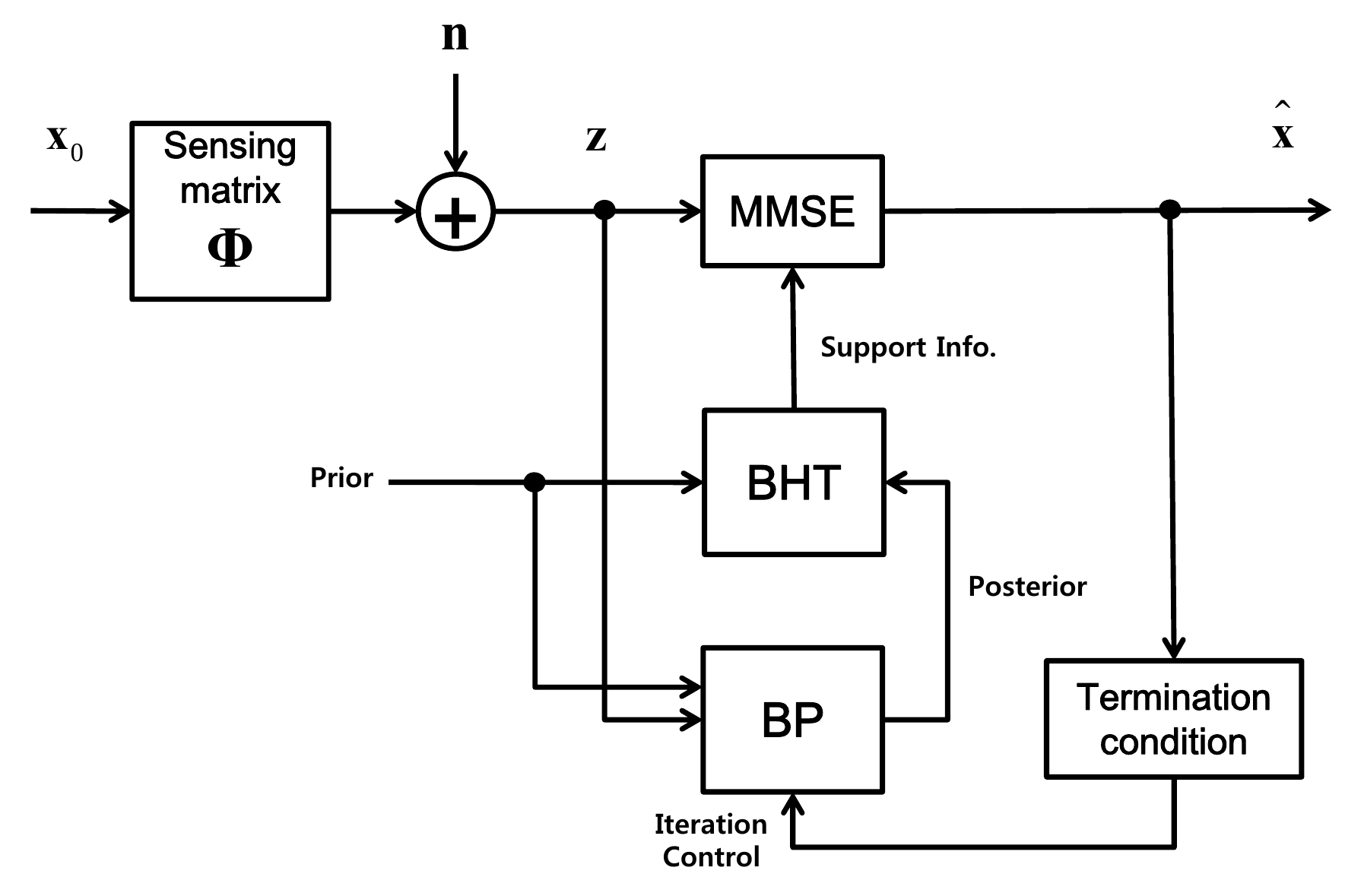}
\caption{System model of CS-BSD.} \label{fig:Fig1-1}
\end{figure}

In this paper, we propose a sparse reconstruction algorithm based on
the Bayesian approach and the use of sparse matrices. We call our
algorithm as \emph{Compressive sensing via bayesian support
detection} (CS-BSD). CS-BSD has the following properties:
\begin{enumerate}
\item Robustness against the measurement noise effects.
\item Ability to perform as the minimum mean square error (MMSE) estimator that has
knowledge of the support set.
\item Fast convergence.
\end{enumerate}
CS-BSD has a \emph{detection-directed} (DD) estimation structure
which consists of signal support detection and signal value
estimation, as shown in Fig.\ref{fig:Fig1-1}. We consider the common
procedure of first using the measurements at hand to detect the
signal's support set. This detected support set is then used in the
model of the sparse signal, and the value estimator is built as if
the detected support set is in fact the correct set. The support
detection component consists of a combination of the \emph{Bayesian
hypothesis test} (BHT) and BP, and signal value estimation using the
detected support set is achieved via an MMSE estimator. CS-BSD
iterates the detection and estimation processes until the constraint
in \eqref{eq:eq_3} is met.

The DD estimation methodology was investigated in \cite{DE_info} for
estimation of noisy signals and have been widely applied to wireless
communication systems \cite{Picchi87},\cite{Godard80}. For CS, the
methodology was first reported in
\cite{HNLEE},\cite{JWKANG_SPARS11}; we tailor the methodology to the
NCS problem by refining that work. The complexity of CS-BSD is $O(N
\log N + KM)$ whereas that of the other BP-based algorithm is $O(N
\log N)$ because CS-BSD includes the cost of MMSE in addition to
that of BP. However, CS-BSD converges faster than the other BP-based
algorithms; thus, its computational cost is lower in practice. In
addition, CS-BSD can be much faster by converting to an parallel
architecture.

The rest of the paper is organized as follows. Section II introduces
the sparse sensing matrix, the prior model for our system model. The
details of  CS-BSD are given in Section III. A few practical issues
are discussed in Section IV. We  compare the numerical results of
CS-BSD to the other recent CS algorithms in Section V. Section VI
concludes the paper.

\section{System Model}

\subsection{Sparse Sensing Matrix $\mathbf{\Phi}$}\label{SBM}
For signal sensing, we employ sparse-Bernoulli matrices
$\mathbf{\Phi}\in\{0,1,-1\}^{M \times N}$, which have been
successfully used in CS recently \cite{CS-BP1}-\cite{BP-SBL}. In the
matrix, sparsely nonzero elements are equiprobably equal to $1$ or
$-1$. We set the sparsity of $\mathbf{\Phi}$ using the fixed column
weight $L$. Because  the column weight rather than the row weight is
fixed, all elements of $\mathbf{x}_0$  have an even chance of being
sensed. In addition, the fixed column weight unifies the energy of
the basis of the measurement space spanned by the column vectors of
$\mathbf{\Phi}$.

\begin{figure}[!b]
\centering
\includegraphics[width=9cm]{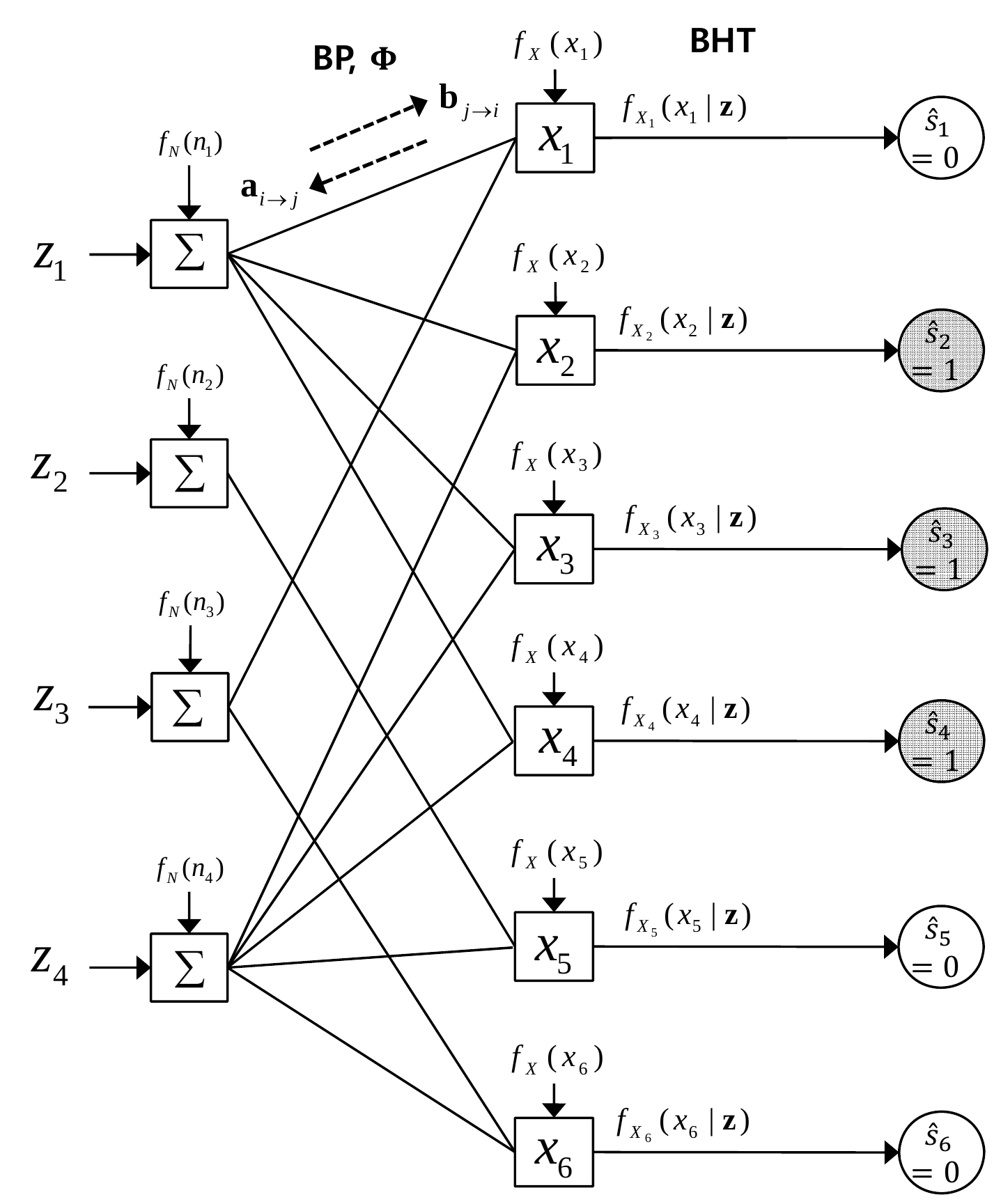}
\caption{Overall flow of support detection in CS-BSD: A case for
$N=6, M=4, L=2$.} \label{fig:Fig4-1}
\end{figure}

With the sparse-Bernoulli matrix, the linear system
$\mathbf{z}=\mathbf{\Phi}\mathbf{x}_0+\mathbf{n}$  can be
represented over a bipartite graph. Let $\mathcal{V}:=\{1,2,...,N\}$
denote a set of indices corresponding to the elements of the signal
vector, $\mathbf{x}_0=[x_{0,1},x_{0,2},...,x_{0,N}]$. Similarly,
$\mathcal{C}:=\{1,2,...,M\}$ denotes a set of indices corresponding
to the elements of the measurement vector,
$\mathbf{z}=[z_1,z_2,...,z_M]$. In addition, we define a set of
edges connecting $\mathcal{V}$ and $\mathcal{C}$ as $\mathcal{E} :=
\{ (i,j) \in \mathcal{V} \times \mathcal{C}\,|\,\,|\phi _{ij}| =
1\}$ where $\phi _{ij}$ is the $(i,j)$-th element of $\mathbf{\Phi
}$. Then, A bipartite graph $\mathcal{G} := (\mathcal{V,C,E})$ fully
describes the neighboring relation in the linear system.
Furthermore, we define the neighbor set of $\mathcal{V}$ and
$\mathcal{C}$ as $N_{\mathcal{V}} (i) := \{ j \in
\mathcal{C}\,|(i,j) \in {\mathcal{E}}\}$ for all $i \in \mathcal{V}$
and $N_\mathcal{C} (j) := \{ i \in \mathcal{V}\,|(i,j) \in
{\mathcal{E}}\}$ for all $j \in \mathcal{C}$, respectively. Note
that $\left|N_{\mathcal{V}} (i)\right|=L$ for all $i \in
\mathcal{V}$ under our assumption regarding $\mathbf{\Phi}$.
Fig.\ref{fig:Fig4-1} depicts a simple example of the graphical
representation corresponding to $N=6, M=4, L=2$.

\subsection{Prior Model}
We limit our discussion to the random vector $\mathbf{x}$ whose
elements are i.i.d. random variables. This assumption is commonly
used in many papers \cite{SBL}-\cite{SuPrEM}. We characterize the
signal sparsity in a probabilistic manner, called \emph{sparsity
rate}.  The sparsity rate $q$ is defined as $q:=\Pr\{x_i \neq 0\}$
for all $i \in \mathcal{V}$. Namely, each signal element
independently belongs to the signal support set with the rate $q$.
The supportiveness of each signal element is represented by a state
variable $s_i$, defined as
\begin{eqnarray}
{s_i} = \left\{ \begin{array}{l}
1,\,\,\,\,\,{\rm{if}}\,\,{x_i} \ne 0\\
0,\,\,\,\,\,{\rm{else}}
\end{array} \right.\text{ for all }\,\, i \in \mathcal{V}.
\end{eqnarray}
Hence, we model the prior density of $x_i$ using a
\emph{spike-and-slab} model originating in a two-state mixture
density as follows:
\begin{eqnarray} \label{eq:eq_4}
f_{x}(x)&:=& qf_{x}( x |s = 1)+ (1 -
q)f_{x}( x |s = 0)\nonumber\\
 &=& q\mathcal{N}(x;0,\sigma _x^2 )+ (1 - q)\delta (x ),
\end{eqnarray}
where $\delta(x)$ indicates a Dirac distribution having nonzero
value between $x \in [0-, 0+]$ and $\int \delta(x)dx =1$. In the
prior density, we use Gaussian density $\mathcal{N}(x;0,\sigma _x^2
)$ for $f_{x}( x |s = 1)$ although it includes the probability mass
at $x_i=0$. The reason is the probability mass at $x_i=0$ is very
small and Gaussian densities are mathematically tractable. In
addition, we drop the index $i$ from the prior density under the
assumption of i.i.d. elements. The spike-and-slab prior has been
widely employed in Bayesian inference problems
\cite{Ishwaran}-\cite{Carvalho} and was recently applied to CS
\cite{TSW-CS} as well.

\section{Proposed Algorithm}
 In this section, we discuss the details of the proposed algorithm
based on the DD estimation structure. The proposed algorithm,
CS-BSD, is an iterative algorithm that repeats the support detection
and  signal value estimation processes until $\mathbf{E} \left\|
{{\bf{\Phi x}} - {\bf{z}}} \right\|_2 \le \epsilon$ is met.

\subsection{Detection of Support Set}
The decoder detects the signal support in each element unit. Namely,
the supportive state of each signal element is detected
independently and converted to the support set information for the
signal. First, the following simple hypothesis test can be
considered for the state detection of $x_i$:
\begin{eqnarray}\label{eq:eq_4-1}
\Pr \{ {x_i} = 0| \mathbf{z} \} \mathop {\mathop \gtrless
\limits_{{H_1}} }\limits^{{H_0}} \Pr \{ {x_i} \neq 0 |\mathbf{z}\}
\text{ for all } i \in \mathcal{V},
\end{eqnarray}
where $H_0$ and $H_1$ are two possible hypotheses. If we marginalize
over $s_i$, the left hand side  of \eqref{eq:eq_4-1} becomes
\begin{eqnarray}\label{eq:eq_4-2}
\begin{array}{l}
\Pr\{x_i = 0 | \mathbf{z} \} = {\sum \limits_{ s_i \in \{ 0,1\} }
{\Pr\{x_i = 0 | \mathbf{z}, s_i \} \Pr\{s_i
|{\bf{z}}\}} } \\
\,\,\,\,\,\,\,\,\,\,\,\,\,\,\,\,\,\,\,\,\,\,\,\,\,\,\,\,\,\,\,\,\,\,\,=
\Pr\{x_i = 0 | \mathbf{z}, s_i =1 \} \Pr\{s_i=1 |{\bf{z}}\}+
 \underbrace{\Pr\{x_i =0 | \mathbf{z}, s_i=0 \}}_{=1}
\Pr\{s_i=0 |{\bf{z}}\}\\
\,\,\,\,\,\,\,\,\,\,\,\,\,\,\,\,\,\,\,\,\,\,\,\,\,\,\,\,\,\,\,\,\,\,\,=
\Pr\{x_i = 0 | \mathbf{z}, s_i =1 \} \Pr\{s_i=1
|{\bf{z}}\}+\Pr\{s_i=0 |{\bf{z}}\},
\end{array}
\end{eqnarray}
where
\begin{eqnarray}
\begin{array}{l}
\Pr \{ {x_i} = 0|\mathbf{z},{s_i} = 0\} = \Pr \{ {x_i} = 0|{s_i} = 0\}\\
\,\,\,\,\,\,\,\,\,\,\,\,\,\,\,\,\,\,\,\,\,\,\,\,\,\,\,\,\,\,\,\,\,\,\,\,\,\,\,\,\,\,\,\,\,\,\,\,\,\,\,\,\,\,
= \int_{  0-}^{  0+} {{f_{x}}({x}|{s} = 0)\,d{x}} \\
\,\,\,\,\,\,\,\,\,\,\,\,\,\,\,\,\,\,\,\,\,\,\,\,\,\,\,\,\,\,\,\,\,\,\,\,\,\,\,\,\,\,\,\,\,\,\,\,\,\,\,\,\,\,
= \int_{  0-}^{  0+} {\delta ({x})\,d{x}}  = 1.
\end{array}
\end{eqnarray}
The right hand side of \eqref{eq:eq_4-1} is
\begin{eqnarray} \label{eq:eq_4-3}
\begin{array}{l}
\Pr\{x_i \neq 0 | \mathbf{z} \} = {\sum \limits_{ s_i \in \{ 0,1\} }
{\Pr\{x_i \neq 0 | \mathbf{z}, s_i \} \Pr\{s_i
|{\bf{z}}\}} } \\
\,\,\,\,\,\,\,\,\,\,\,\,\,\,\,\,\,\,\,\,\,\,\,\,\,\,\,\,\,\,\,\,\,\,\,=
\Pr\{x_i \neq 0 | \mathbf{z}, s_i =1 \} \Pr\{s_i=1 |{\bf{z}}\}+
\underbrace { \Pr\{x_i \neq 0 | \mathbf{z}, s_i=0 \}}_{=0}
\Pr\{s_i=0 |{\bf{z}}\}\\
\,\,\,\,\,\,\,\,\,\,\,\,\,\,\,\,\,\,\,\,\,\,\,\,\,\,\,\,\,\,\,\,\,\,\,=
\Pr\{x_i \neq 0 | \mathbf{z}, s_i =1 \} \Pr\{s_i=1 |{\bf{z}}\},
\end{array}
\end{eqnarray}
where
\begin{eqnarray}
\Pr\{x_i \neq 0 | \mathbf{z}, s_i=0 \} =\Pr\{x_i \neq 0 | s_i=0 \}\\
=\int \limits_{\mathbb{R}/\{0\}} {\delta ({x})\,d{x}}  = 0.
\end{eqnarray}
From \eqref{eq:eq_4-2} and \eqref{eq:eq_4-3}, the hypothesis test in
\eqref{eq:eq_4-1} is refined as
\begin{eqnarray} \label{eq:eq_5}
\begin{array}{l}
 \frac{{\Pr \{ s_i  = 0|{\bf{z}}\} }}{{\Pr \{ s_i =
1|{\bf{z}}\} }} \mathop {\mathop \gtrless \limits_{{H_1}}
}\limits^{{H_0}} \Pr\{x_i \neq 0 | \mathbf{z}, s_i =1 \}-\Pr\{ x_i =
0|{\bf{z}},s_i  = 1\}\\
\,\,\,\,\,\,\,\,\,\,\,\,\,\,\,\,\,\,\,\,\,\,\,\,\,\,\,\,\,\,\,\,\,\,=1-
2\times \Pr\{ x_i = 0|{\bf{z}},s_i  = 1\}.
\end{array}
\end{eqnarray}
Here
\begin{eqnarray}
\Pr \{ {x_i} = 0|\mathbf{z},{s_i} = 1\} = \int_{  0-}^{  0+}
{{f_{x_i}}({x}|\mathbf{z},{s_i} = 1)\,d{x}},
\end{eqnarray}
where the posterior density, $f_{x_i}({x}|\mathbf{z},{s_i} = 1)$, is
Gaussian because the signal and noise are assumed to be Gaussian
(see p.326 in \cite{Kay}). The term of $\Pr\{ x_i = 0|{\bf{z}},s_i =
1\}$ in right hand side of \eqref{eq:eq_5} is caused by the use of
Gaussian density $\mathcal{N}(x;0,\sigma _x^2 )$ for the prior of
nonzero $x_i$. Because the variance of $f_{x_i}({x}|\mathbf{z},{s_i}
= 1)$ is a function of the noise variance, the probability $\Pr \{
{x_i} = 0|\mathbf{z},{s_i} = 1\}$ is very small, and it approaches
zero as the SNR increases. Therefore, we suggest setting the
threshold of the hypothesis test in \eqref{eq:eq_5} to 1. This
implies that the hypothesis test can detect the supportive state of
the signal elements with a high success probability if SNR is
sufficiently high.

We now describe how to obtain the probability ratio, $ \frac{{\Pr \{
s_i  = 0|{\bf{z}}\} }}{{\Pr \{ s_i = 1|{\bf{z}}\} }} $. By
factorizing over $x_i$, the ratio becomes
\begin{eqnarray}\label{eq:eq_6}
{\frac{{\Pr \{ s_i  = 0|{\bf{z}}\} }}{{\Pr \{ s_i  = 1|{\bf{z}}\}
}}}=  {\frac{{\int  {\Pr \{ s_i  = 0|{\bf{z}},x_i \} f_{x_i}(x
|\mathbf{z} ) }dx }}{{\int  {\Pr \{ s_i = 1|{\bf{z}},x_i \} f_{x_i}(
x |\mathbf{z} ) }dx }}} \mathop {\mathop \gtrless \limits_{{H_1}}
}\limits^{{H_0}} 1,
\end{eqnarray}
where $f_{x_i}( x |\mathbf{z} )$ denotes the posterior density of
$x_i$ given $\mathbf{z}$. The signal elements are not i.i.d. anymore
given $\mathbf{z}$. In \eqref{eq:eq_6}, $\Pr \{ s_i |{\bf{z}},x_i \}
= \Pr \{s_i |x_i \}$ holds true since the measurements $\mathbf{z}$
does not provide any additional information on the state given
$x_i$. Using the Bayesian rule and the prior information, we finally
obtain the hypothesis test as the following form:
\begin{eqnarray}\label{eq:eq_7}
{\frac{{\Pr \{ s_i  = 0|{\bf{z}}\} }}{{\Pr \{ s_i  = 1|{\bf{z}}\}
}}}={\frac{{\int  {\frac{{f_{x}(x |s = 0) \Pr \{ s = 0\} }}{{f_{x}(x
) }}f_{x_i}(x |\mathbf{z} )}dx }}{{\int {\frac{{f_{x}( x |s = 1) \Pr
\{ s = 1\} }}{{f_{x}( x ) }}f_{x_i}(x |\mathbf{z} ) }dx }}} \mathop
{\mathop \gtrless \limits_{{H_1}} }\limits^{{H_0}}  1.
\end{eqnarray}
Since we know the prior of the state $Pr\{s\}$ from the sparsity
rate, \emph{i.e.}, $\Pr \{ s = 1\}=q$, we can move the prior term to
the right side, and then treat it as a threshold for the hypothesis
test $\gamma:={\frac{{\Pr \{ s = 1\} }}{{\Pr \{ s = 0\}
}}}={\frac{q}{(1-q)}}$. Therefore, the  state of each elements can
be sensed from the corresponding posterior and prior densities.

\bigskip
\newtheorem{def1}{\textbf{Definition}}
\begin{def1}[BHT for state detection]
Let $\widehat{s}_i$ denote the detected state of $x_i$;
$f_{x_i}(x|\mathbf{z} )$ indicates the posterior density of $x_i$;,
and $f_{x}(x|s)$ denotes the conditional prior density of a signal
element  given the state. Then, state detection for all $i \in
\mathcal{V}$ is performed by choosing the hypothesis that result
from
\begin{eqnarray}\label{eq:def1}
{\frac{{\Pr \{ s_i  = 0|{\bf{z}}\} }}{{\Pr \{ s_i  = 1|{\bf{z}}\}
}}}={\frac{{\int  {\frac{{f_{x}(x |s = 0) }}{{f_{x}(x ) }}f_{x_i}(x
|\mathbf{z} )}dx }}{{\int {\frac{{f_{x}( x |s = 1) }}{{f_{x}( x )
}}f_{x_i}(x |\mathbf{z} ) }dx }}} \mathop {\mathop \gtrless
\limits_{{H_1}} }\limits^{{H_0}}  \gamma,
 \end{eqnarray}
\begin{eqnarray}
\text{where }\left\{ \begin{array}{l}
{H_0}:\widehat s_i = 0\\
{H_1}:\widehat s_i = 1
\end{array} \right.,\,\,\,\gamma:=q/(1-q).
\end{eqnarray}
\end{def1}

\subsection{Belief Propagation for Posterior Update}
The posterior density used for the BHT is obtained and updated at
every iteration via BP. Our BP process is similar to that in
\cite{CS-BP1},\cite{CS-BP2} and was independently devised from
\cite{BP-SBL},\cite{SuPrEM}. Distinctively, our BP process uses the
information on the noise distributions $f_{n_j}(n) =
\mathcal{N}(n;0,\sigma^2_n)$  under the i.i.d. zero-mean Gaussian
noise assumption.

Using Bayesian rule, we can represent the posterior density of $x_i$
in the form of $ {\rm{Posterior = Prior}} \times
\frac{{{\rm{Likelihood}}}}{{{\rm{Evidence}}}}$, given as
\begin{eqnarray}\label{eq:eq_12}
f_{x_i}(x|\mathbf{z}) =   f_{x}(x)\times \frac{ { f_{\mathbf{z}}(
{{\bf{z}}|x_i)}}}{f_{\mathbf{z}}(\mathbf{z})}.
\end{eqnarray}
If the sensing matrix $\mathbf{\Phi}$ is sufficiently sparse such
that the corresponding bipartite graph is tree-like, we postulate
that the elements of $\mathbf{z}$ associated with $x_i$ are
independent of each other given $x_i$ \cite{Richardson}. Under the
tree-like assumption, we can decompose the likelihood density
$f_{\mathbf{z}}( {\bf{z}}|x_i)$ to the product of densities:
\begin{eqnarray} \label{eq:eq_13}
f_{x_i}(x|\mathbf{z}) \propto f_{x}(x) \times \prod\limits_{j \in
N(i)} f_{z_j}({z}|{x_i}).
\end{eqnarray}
We call each decomposition of the likelihood, $f_{z_j}(z|x_i)$ the
\emph{measurement density}. Theorem 1 below demonstrates that the
measurement density can be composed of the densities of the
associated signal elements.

\bigskip
\newtheorem{th1}{\textbf{Theorem}}
\begin{th1}[Measurement density in BP] The measurement density
$f_{z_j}(z|x_i)$ is expressed as the linear convolution of  all the
associated distributions of the signal elements and the
corresponding noise distribution $f_{n_j}(n)$ as follows:
\begin{eqnarray} \label{eq:th1}
f_{z_j}(z|x_i)= \delta(z - z_j) \otimes f_{n_j}(n) \otimes
\left(\bigotimes\limits_{k \in N_{\mathcal{C}}(j)\backslash \{i\} }
{f_{x_k}}(x) \right),
\end{eqnarray}
$\text{ for all } (i,j) \in \mathcal{E}$, where   $\otimes$ and
$\bigotimes$ are the operator for linear convolution and the linear
convolution of a sequence of functions, respectively\\
\emph{Proof}: See Appendix A.
\end{th1}
\bigskip

Therefore, the essence of the BP-process is to update the signal and
 measurement densities by exchanging probability density messages,
associated with the neighboring relation in the bipartite graph. Let
$\mathbf{a}_{i \rightarrow j}$ denote the message from the $i$-th
signal element to the $j$-th measurement element, called the signal
message; $\mathbf{b}_{j \rightarrow i}$ is the message from the
$j$-th measurement element to the $i$-th signal element, called the
measurement message. The signal message is an approximation of the
density of the signal element, \emph{i.e.}, $\mathbf{a}_{i
\rightarrow j} \approx f_{x_i}(x|\mathbf{z})$ and it is obtained
from \eqref{eq:eq_13}  simply by replacing the measurement density
with the measurement message of the previous iteration. Note that in
BP-process the message coming from the $j$-th measurement is
excluded in the calculation of $\mathbf{a}_{i \rightarrow j}$. Thus,
the signal message at the $l$-th iteration is expressed as
\begin{eqnarray}\label{eq:eq_14}
\mathbf{a}_{i \rightarrow j}^l:=\eta\left[ {{f_{x}}(x)  \times
\prod\limits_{k \in N_{\mathcal{V} }(i)\backslash\{j\}} {{\bf{b}}_{k
\rightarrow i}^{l-1} } } \right]
\end{eqnarray}
$\text{ for all } (i,j) \in \mathcal{E}$, where  $\eta[\cdot]$ is
the normalization function to make $ \int{\mathbf{a}_{i\rightarrow
j} }dx=1$. Similarly, the measurement message approximates the
measurement density, \emph{i.e.}, $\mathbf{b}_{j \rightarrow i}
\approx f_{{z_j}}({z}|{x_i})$, and it is obtained from the
expression of \eqref{eq:th1} by replacing the associated densities
of signal elements $f_{x_k}(x)$ with the signal messages for the
purpose of iteration , that is,
\begin{eqnarray} \label{eq:th1-1}
\mathbf{b}_{j \rightarrow i}^l:= \delta(z - z_j) \otimes f_{n_j}(n)
\otimes \left(\bigotimes \limits_{k \in N_{\mathcal{C}}(j)\backslash
\{i\} } {\mathbf{a}_{k \rightarrow j}^{l}} \right).
\end{eqnarray}

\begin{figure}[!t]
\centering
\includegraphics[width=10cm]{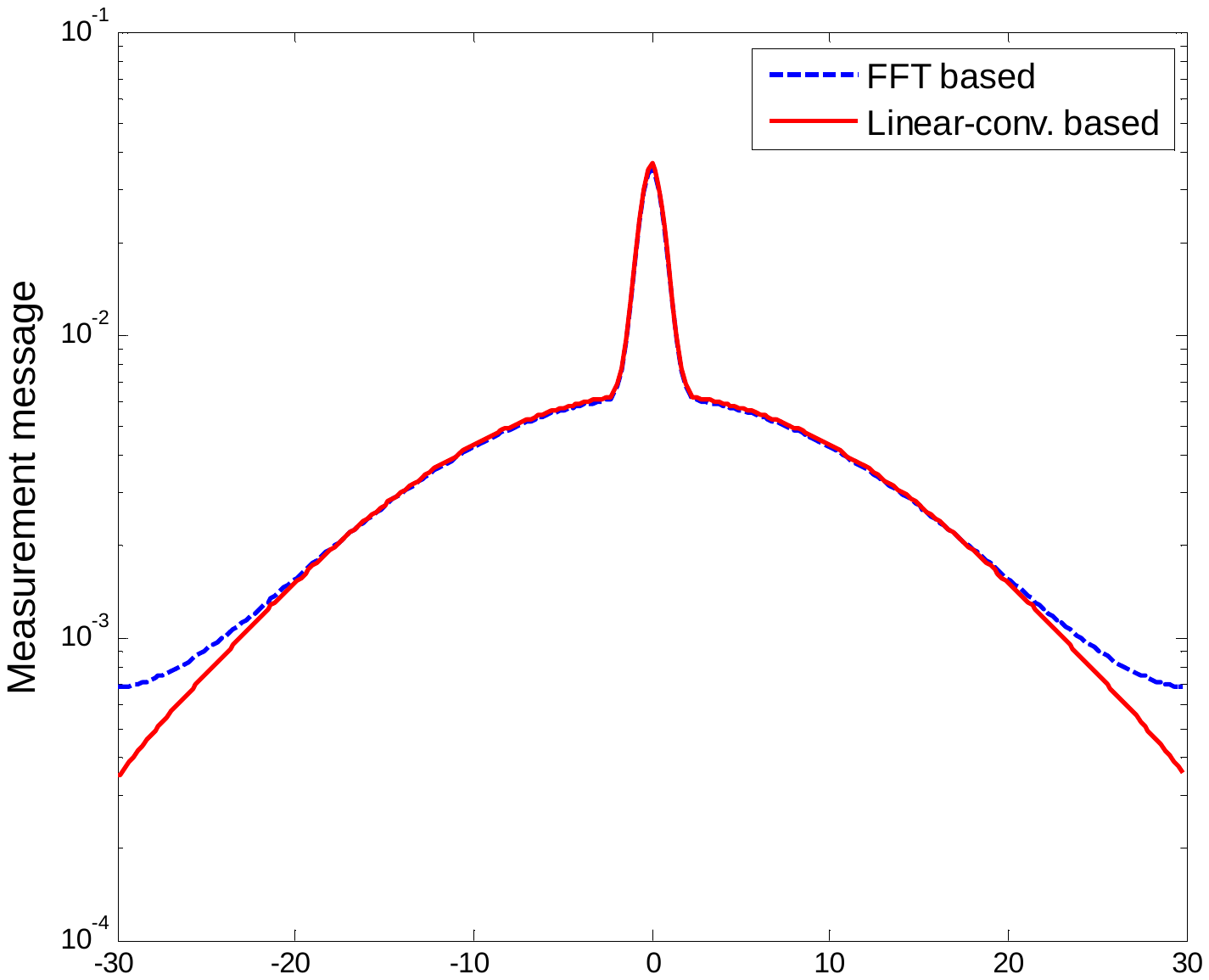}
\caption{Calculation gap between use of linear convolution and
FFT-based convolution in measurement message calculation.}
\label{fig:Fig4-3}
\end{figure}

The convolution operations in \eqref{eq:th1-1} can be efficiently
computed by using the \emph{Fast fourier transform} (FFT).
Therefore, we express for the measurement message calculation as
\begin{eqnarray}\label{eq:eq_16}
{\bf{b}}_{j \to i}^l : = {\bf{F}}^{ - 1} \left[ {{\bf{F}}\delta (z -
z_j ) \times {\bf{F}}f_{n_j}(n) \times \prod\limits_{k \in
N_{\mathcal{C}} (j)\backslash \{ i\} } {{\bf{Fa}}_{k \to j}^l } }
\right]
\end{eqnarray}
where $\mathbf{F} \in \mathbb{C}^{N_d \times N_d}$ denotes a Fourier
matrix of size $N_d$. In fact, the use of the FFT brings a small
calculation gap between this result and that of \eqref{eq:th1} since
the FFT-based calculation performs a circular convolution that
produces output having a heavy tail, as shown in
Fig.\ref{fig:Fig4-3}. The heaviness increases as the corresponding
row weights in $\mathbf{\Phi}$ increase. However, the difference is
can be ignored, especially when the densities are bell-shaped
distributions.

Finally, the update of the posterior density of $x_i$ at the $l$-th
iteration is provided as given in Definition 2.
\bigskip
\newtheorem{def2}[def1]{\textbf{Definition}}
\begin{def2}[Posterior update in BP]\label{def2}
Let $\mathbf{b}_{j \rightarrow i}^l$ denote a measurement message at
the $l$-th iteration for all $(i,j) \in \mathcal{E}$. Then, the
posterior density of $x_i$ at the $l$-th iteration is calculated by
\begin{eqnarray} \label{eq:eq_17}
f_{x_i^l}(x |\mathbf{z} ) =\eta\left[ { f_{x}(x) \times
\prod\limits_{j \in N_{\mathcal{V}} (i)} {{\mathbf{b}_{j \rightarrow
i}^{l}} } } \right],
\end{eqnarray}
where  $\eta[\cdot]$ is the normalization function that makes $
\int{f_{x_i^l}(x |\mathbf{z} )  }dx=1$.
\end{def2}

\subsection{Detection-Directed Estimation of Signal Values}
We now describe signal value estimation based on the DD estimation
structure. The DD estimator is basically an estimator that
determines how to act on the input data directed by the information
from the detector. In CS-BSD, the detector provides the support
information $\widehat{\mathbf{s}}^l$, and the value estimator then
finds the signal values as if the detected support set is the
correct set at each iteration. That is,
\begin{eqnarray}\label{eq:eq_18}
{\widehat {\bf{x}}^l} =\arg \mathop {\max }\limits_{\bf{x}}
{f_{\bf{x}}}({\bf{x}}|{\bf{z}},\,{\bf{s}} = {\widehat
{\bf{s}}}^l)\,\,\,\,\,\,s.t.\,\, \mathbf{E} \left\| { ({\bf{\Phi x}}
- {\bf{z}})} \right\|_2  \le \epsilon,
\end{eqnarray}
where the estimator decides that $\widehat{x}_i^l =0$ for all $i\in
\mathcal{V}:\widehat{s}_i^l =0$. From the argument in
\eqref{eq:eq_5}, the DD estimate converges to the true signal
$\mathbf{x}_0$ since the detected support set becomes the correct
set as SNR and the number of iterations $l$ increases. This DD
methodology makes no general claim regarding optimality of the
solution; however, it is common and often successful. Let
$\mathbf{x}_{supp}^l \in \mathbb{R}^{||\widehat{ \mathbf{s}}^l||_0}$
denote a random vector consisting of the elements with
$\widehat{s}_i^l=1$. Then, the problem in \eqref{eq:eq_18} is
reduced to
\begin{eqnarray}\label{eq:eq_19}
\begin{array}{l}
\widehat {\bf{x}}_{{supp}}^l = \arg \mathop {\max }\limits_{ {\bf{x}} } {f_{{\bf{x}}_{{supp}}^l}}({\bf{x}}|{\bf{z}},\,\,{\bf{s}} = {\widehat {\bf{s}}}^l)\\
= \arg \mathop {\max }\limits_{ {\bf{x}} }
{f_{\bf{z}}}({\bf{z}}|{\bf{x}}_{{supp}}^l,\,{\bf{s}} = {\widehat
{\bf{s}}}^l){f_{{\bf{x}}_{{supp}}^l}}({\bf{x}}|{\bf{s}} = {\widehat
{\bf{s}}}^l).
\end{array}
\end{eqnarray}
Since $\mathbf{x}_{supp}^l$ and the noise elements are assumed to be
zero-mean i.i.d. Gaussian random variables with variance
$\sigma_x^2$ and $\sigma_n^2$ respectively, the MAP estimation in
\eqref{eq:eq_19} is recast as
\begin{eqnarray}
\begin{array}{l}
{\widehat {\bf{x}}_{supp}}^l = \arg \mathop {\min
}\limits_{{\bf{x}}_{supp}^l} \frac{1}{{\sigma _n^2}}\left\|
{{\bf{z}} - {\bf{\Phi }}_{supp}^l{\bf{x}}_{supp}^l} \right\|_2^2 +
\frac{1}{{\sigma _x^2}}\left\| {{\bf{x}}_{supp}^l} \right\|_2^2,
\end{array}
\end{eqnarray}
where $\mathbf{\Phi}_{supp}^l$ denotes a submatrix of
$\mathbf{\Phi}$ corresponding to $i \in \mathcal{V}:
\widehat{s}_i=1$. In addition, the MAP and MMSE estimates are
identical, assuming the signal and noise are Gaussian (see p.358 in
\cite{Kay}). Therefore, the estimate $\widehat{\mathbf{x}}_{supp}^l$
can be obtained by the MMSE estimator
\begin{eqnarray}\label{eq:eq_MMSE}
\widehat{\bf{x}}_{supp}^l  = \left(\frac{1}{{\sigma _x^2}}{\bf{I}} +
\frac{1}{{\sigma _n^2}} { {\bf{\Phi }}_{supp}^{l^{\,\,*} } {\bf{\Phi
}}_{supp}^l } \right)^{ - 1} \frac{1}{{\sigma _n^2}}{\bf{\Phi
}}_{supp}^{l^{\,\,*} } {\bf{z}}.
\end{eqnarray}
To combine the support information $\widehat{\mathbf{s}}^l$ and the
estimated values $\widehat{\bf{x}}_{supp}^l$, we define an index set
$\mathcal{U}^l:=\{1,...,{\left\| {\widehat {\bf{s}}^l}
\right\|_0}\}$ corresponding to the elements
$\mathbf{x}_{supp}^l=[x_{supp,1},...,x_{supp,{\left\| {\widehat
{\bf{s}}^l} \right\|_0}}]$ and a bijective mapping function $h: \{ i
\in \mathcal{V}|\widehat{s}_i^l =1\} \rightarrow \mathcal{U}^l$.
Then, the reconstruction at each iteration is readily obtained from
\begin{eqnarray}
\widehat{x}_i^l  = \left\{ \begin{array}{l}
\widehat{x}_{supp,h(i)}^l  ,\,\,\,\,{\text{if }} \widehat{s}_i^l =1 \\
 0,\,\,\,\,\,\,\,\,\,\,\,\,\,\,\,\,\,\,\,\,\,\,\,\,{\text{o.w.}}\\
 \end{array} \right.
\end{eqnarray}
for all $i \in \mathcal{V}$. CS-BSD is summarized in Algorithm
\ref{alg:CS-BSD}.

\section{Practical issues}
\subsection{Complexity}
We implement the BP process in CS-BSD based on the sampled-message
approach in \cite{CS-BP2}. The density messages $\mathbf{a}_{ i
\rightarrow j}, \mathbf{b}_{ j \rightarrow i}$ are vectors of size
$N_d$ where $N_d$ is chosen to be power of two for efficient use of
FFT. Next, we analyze the complexity of CS-BSD by considering each
part seperately.

\subsubsection{Support detection} Let us consider the complexity of
BP first. Since the matrix $\Phi$ has the fixed column weight $L$
and the size for a density vector is $N_d$, the decoder requires
$O(LN_d)$ flops per iteration to  calculate the signal message
$\mathbf{a}_{i \rightarrow j}$ in \eqref{eq:eq_14}, and  $O(
{\frac{{NLN_d }}{M}\log N_d })$ flops per iteration to calculate the
measurement message $\mathbf{b}_{j \rightarrow i}$ in
\eqref{eq:th1}, since the row weight is $NL/M$ on average and the
cost of the FFT-based convolution is $O(N_d\log N_d)$. Hence, the
per-iteration cost for all probability messages is
$O(NLN_d+M{\frac{{NLN_d }}{M}\log N_d })$ flops. For the BHT in
\eqref{eq:def1},  the decoder requires $O(N_d)$ flops to calculate a
likelihood ratio. The cost for the hypothesis test is much smaller
than that of BP; therefore, it is ignored.

\subsubsection{Signal value estimation}
Let us fix the signal sparsity as the expected value of the
cardinality of the support set, \emph{i.e.}, $K:=E[{\left\|
{{{\bf{x}}}} \right\|_0}]=Nq$, for  purpose of comparison. Then, the
complexity of the MMSE estimation in \eqref{eq:eq_MMSE} depends
strongly upon $K$ such that conventionally it requires $O(KM)$ flops
if QR decomposition is used \cite{Bjorck}. Thus, the total
complexity of CS-BSD is $O\left( N_{iter} NLN_d \log N_d + N_{iter}
KM \right)$ flops where $N_{iter}$ denotes the number of iterations.
If  $L$ and $N_d$ are fixed, the complexity of CS-BSD can be
simplified to $O(N_{iter}N+N_{iter}KM)$ flops.  The BP process is
known to converge within $N_{iter}=O(\log N)$ \cite{Mackey} such
that its complexity is $O( N\log N +KM\log N)$. If we fix the number
of iteration $N_{iter}$ empirically, we can remove the MMSE
operation from the iterations. In that case, the complexity is
reduced to $O( N\log N +KM)$.

\subsection{Parallelization of Belief Propagation}
The BP process for finding the posterior finding can be implemented
using a parallel architecture. Indeed, many parallelized BP
algorithms, with applications to LDPC codes, have demonstrated
superior performance in \cite{Howland}-\cite{Wang}. The graph
representation of the sparse sensing matrix shows that the
dependencies of the message calculations for any signal elements (or
measurement elements) depend only upon the corresponding measurement
elements (or signal elements). This allows all messages in BP to be
computed in a parallel manner. Therefore, implementing BP on a
parallel architecture for BP yields low power consumption,
high-speed decoding, and simple logic \cite{Howland}.

\section{Numerical results} \label{Numresult}
We demonstrate the advantages of CS-BSD using simulation results in
several different settings. To show its average performance, we take
200 Monte Carlo trials for each point in the simulation. In each
trial, we generate the deterministic sparse signal $\mathbf{x}_0$
with $N=1024$ and $\sigma_x=10$ whose values are represented with
finite precision. The finite precision is provided by 6-bit
quantization such that each signal value has 64 levels. This
assumption of finite precision for the signal values is reasonable
in terms of digital signal processing and implementation. In
addition, we restrict the magnitude level of the signal elements to
$|x_{i}| \leq 3\sigma_x$ for the same reason. We define the SNR as
\begin{eqnarray}
\text{SNR :} =10 \log_{10} \frac{{E{\left\| {{\bf{\Phi}} \mathbf{x}}
\right\|_2^2 } }}{{M\sigma _{n }^2 }} \text{ dB }
\end{eqnarray}
and $M/N$ as the undersampling ratio for signal acquisition.

\subsection{SER Performance of Support Detector}
\begin{figure}[!b]
\centering
\includegraphics[width=15cm]{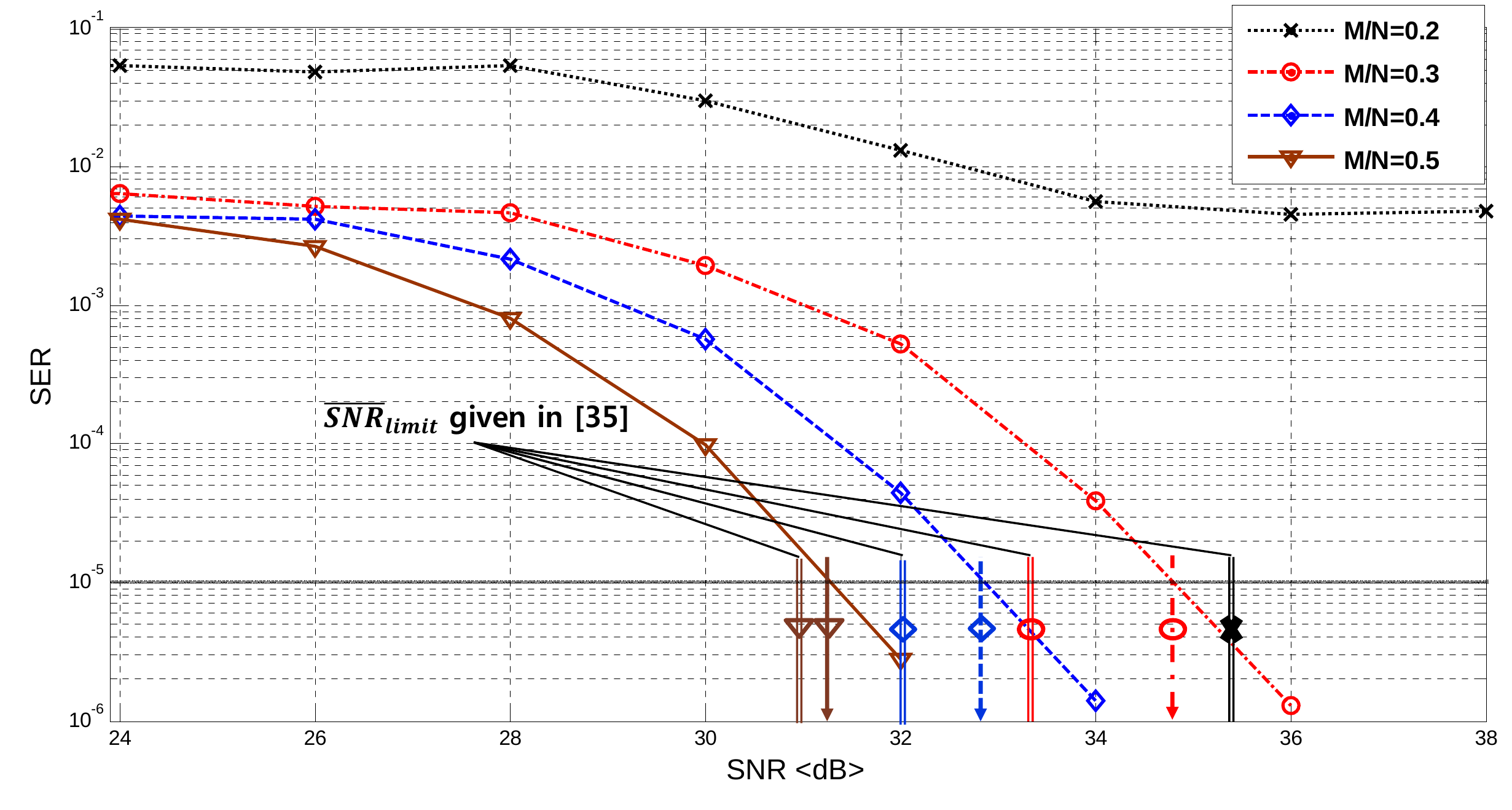}
\caption{SER for support detection of CS-BSD over SNR for $N=1024,
q=0.05$, $L=4$, and $N_d=64$. The double-lines indicate
$\overline{\text{SNR}}_{limit}$ and the downarrow-lines denote the
SNR threshold of the support detector.}\label{fig:Fig5-1}
\end{figure}

To determine the performance of the support detector in CS-BSD, we
defined the \emph{state error rate} (SER) as:
\begin{eqnarray}
{\text{SER}}: =avg \left[ \frac{{\# \{ i \in \mathcal{V}
|\widehat{s}_i \ne s_{0,i}\} }}{N}\right],
\end{eqnarray}
where $s_{0,i}$ is the state variable corresponding to the true
signal value $x_{0,i}$. We simulate the SER performance as a
function of the SNR for a variety of undersampling ratio $M/N$. In
this simulation, we set $q=0.05$, $N_d=64$, and $L=4$. In addition.
we compare the SER performance to a theoretical limit on the support
recovery given by Fletcher \emph{et al.} \cite{Fletcher}. They found
a necessary condition for \emph{maximum-likelihood} (ML) estimation
to asymptotically recover the support set if the sensing matrix has
i.i.d. Gaussian entries. The ML estimation is described as
\begin{eqnarray}
\mathop {\max }\limits_{\mathcal{J}} ||{\bf{P}}_\mathcal{J}
\mathbf{z}||^2_2 \,\,\,\,\,\,s.t.\,\,| \mathcal{J} |={\left\|
{{{\bf{x}}}} \right\|_0},
\end{eqnarray}
where the signal sparsity ${\left\| {{{\bf{x}}}} \right\|_0}$ is
assumed to be known, $\mathcal{J} \subseteq \mathcal{V}$ is a subset
of the index set of the signal, and ${\bf{P}}_\mathcal{J}
\mathbf{z}$ denotes the orthogonal projection of  $\mathbf{z}$ onto
the subspace spanned by columns of $\mathbf{\Phi}$ corresponding to
$\mathcal{J}$. Namely, the ML estimate is a subset of $\mathcal{V}$
such that the subspace spanned by the corresponding columns of
$\mathbf{\Phi}$ contain the maximum energy of $\mathbf{z}$. We
rewrite the necessary condition in terms of SNR such that
\begin{eqnarray}
\text{SNR}>{\text{SNR}}_{limit}:=
 \frac{2\times {\left\| {{{\bf{x}}}}
\right\|_0}\log (N -{\left\| {{{\bf{x}}}} \right\|_0}) }{{(M -
{\left\| {{{\bf{x}}}} \right\|_0} + 1)\times \text{MAR}}},
\end{eqnarray}
where minimum-to-average ratio (MAR) is defined as
\begin{eqnarray}
\text{MAR}: = \frac{{\mathop {\min }\limits_{j:x_{j}\neq 0}
|{x_{j}}{|^2}}}{{\left\| {\bf{x}} \right\|_2^2/{\left\| {\bf{x}}
\right\|_0} }}.
\end{eqnarray}
In this comparison, we used 200 Monte Carlo trials to find the
average SNR$_{limit}$, \emph{i.e.},
$\overline{\text{SNR}}_{limit}:=avg[\text{SNR}_{limit}]$. In
Fig.\ref{fig:Fig5-1}, the SER curves show a waterfall behavior; the
curves decline rapidly  to less than $10^{-5}$ beyond a certain
threshold SNR. This behavior supports the argument in
\eqref{eq:eq_5} that the BHT achieves successful support detection
in the high SNR regime. We consider the SER=$10^{-5}$ bound as an
almost error-free bound since it is much less than the rate of one
state error $1/N\approx 10^{-3}$ when $N=1024$. The threshold SNR
for the error-free bound is roughly 34.8 dB for $M/N$=0.3, 32.9 dB
for $M/N=0.4$, and 31.1 dB for $M/N=0.5$. Remarkably, this threshold
SNR approaches to$\overline{\text{SNR}}_{limit}$ as $M/N$ increases.
For example, the gap between the limit and the simulation result is
0.58 dB for $M/N=0.3$; however, the gap is only 0.2 dB for
$M/N=0.5$. For $M/N=0.2$, since the sensing matrix $\mathbf{\Phi}$
is not sufficiently sparse, the tree-like assumption regarding
$\mathbf{\Phi}$ is rarely satisfied. Such a fact occasionally causes
the BP-process to diverge, leading to severe errors in support
detection.

\subsection{MSE Performance Comparison}
We consider the reconstruction performance in terms of normalized
\emph{means square error} (MSE), which is defined as
\begin{eqnarray}
{\text{MSE}}: =avg \left[ \frac{ {\left\| {\widehat{\bf{x}} -
{\bf{x}}_0} \right\|_2^2 } }{{\left\| {\bf{x}}_0 \right\|_2^2
}}\right].
\end{eqnarray}
 We compare our algorithm to several recent CS reconstruction
algorithms: 1) CS-BP \cite{CS-BP1},\cite{CS-BP2}, 2) L1-DS via
linear programming \cite{candes2}, 3) Bayesian CS (BCS) \cite{BCS},
4) CoSaMP \cite{Cosamp}, and 5) SuPrEM (reweighted version)
\cite{SuPrEM}. For BCS and SuPrEM, we obtained the source code from
each author's webpage; for CoSaMP we used Stephen Becker's code
(available at
\url{http://www.ugcs.caltech.edu/~srbecker/algorithms.shtml}). L1-DS
is provided by the L1-MAGIC package (available at
\url{http://users.ece.gatech.edu/~justin/l1magic/}).We implemented
CS-BP algorithm by using the sampled-message approach and upgrading
the original algorithm to use the noise information. For CS-BP, we
used the sparse-Bernoulli sensing matrix with $L=4$; for SuPrEM, we
use a sensing matrix generated from a low-density frame
\cite{SuPrEM} with the same parameters ($N$, $M$, $L$). L1-DS,
CoSaMP and BCS were used with a Gaussian sensing matrix having the
same column energy as the sparse-Bernoulli matrix, for fairness,
\emph{i.e.}, $\left\| {\mathbf{\phi}
_{j,Gaussian}}\right\|_2^2=\left\|{\mathbf{\phi}_{j,Sparse}}\right\|_2^2=L$.
The sparsity of an input parameter in CoSaMP and SuPrEM was set
according to the expectation of the cardinality of the support set
$K:=E[{\left\| \mathbf{x} \right\|_0}]=Nq$. Those algorithms are
summarized in Table \ref{table1}, with respect to thier complexity,
type of  sensing matrix, prior type, and algorithm type.
\subsubsection{Comparison with respect to SNR}
\begin{figure}[!t]
\centering
\includegraphics[width=15cm]{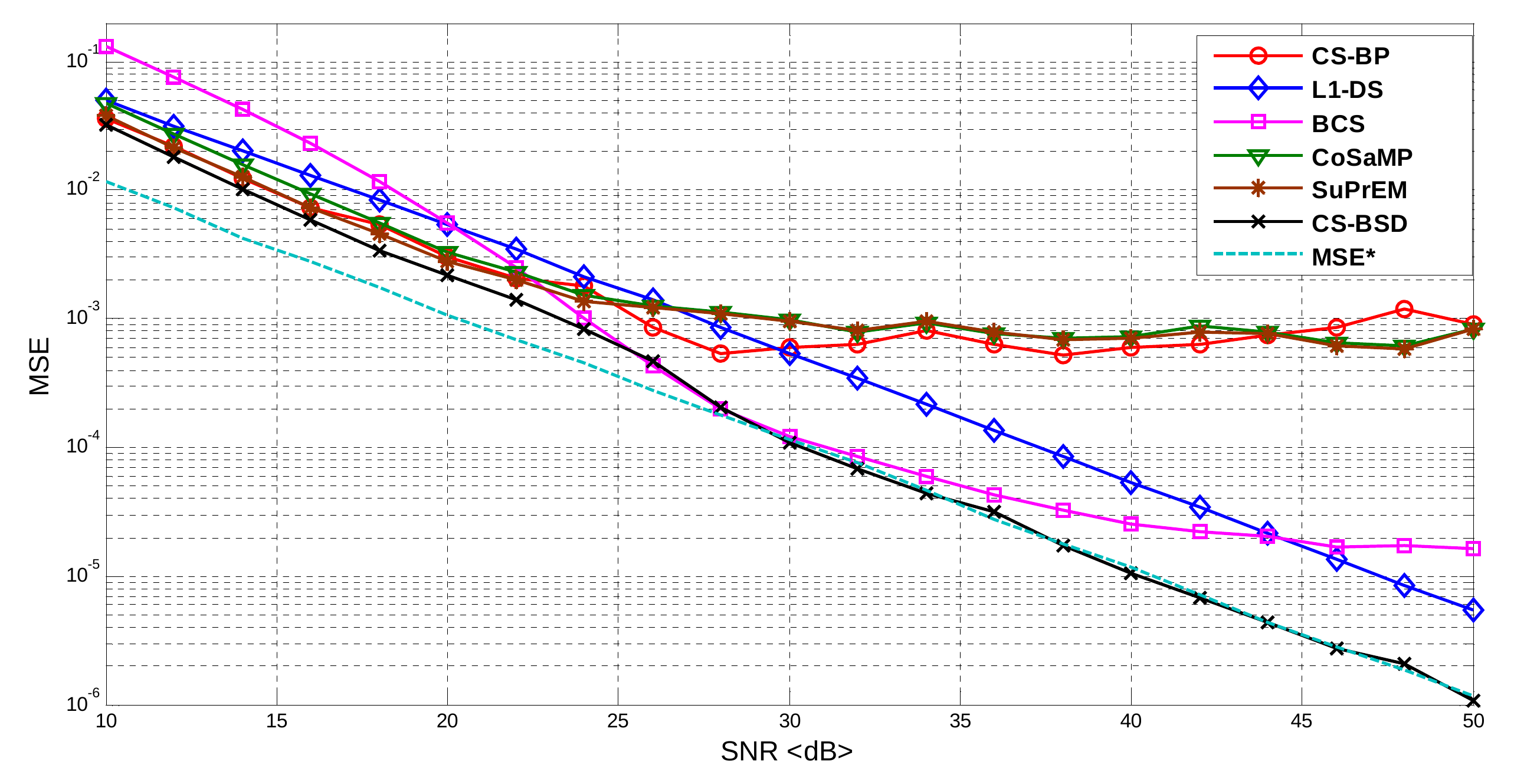}
\caption{MSE comparison over SNR for $N=1024$, $q=0.05$, $M/N=0.5$,
and $N_d=64$ where \FCR{MSE$^*$} denotes the MSE of the MMSE
estimator which has the support knowledge.}\label{fig:Fig5-2}
\end{figure}

In Fig.\ref{fig:Fig5-2}, we show the MSE performance as a function
of SNR where $M/N=0.5$, $q=0.05$, and $N_d=64$. In the high SNR
regime, the advantage of CS-BSD becomes remarkable. As the SNR
increases, the MSE of CS-BSD approaches to that of an MMSE estimator
that has knowledge of the support set, defined as
\begin{eqnarray}
{\text{MSE}}^* := \frac{{{\rm{Tr}} \left[ { \left(\frac{1}{{\sigma
_x^2}}{\bf{I}} + \frac{1}{{\sigma _n^2}}{{\bf{\Phi }}_{supp}^*
{\bf{\Phi }}_{supp} } \right)^{ - 1} } \right]}}{{\left\|
{{\bf{x}}_{{{0,supp}}} } \right\|_2^2 }},
\end{eqnarray}
where Tr$[\cdot]$ denote the matrix trace operation. Beyond SNR=31
dB, since the SER of CS-BSD is almost error-free, the MSE
performance achieves ${\text{MSE}}^*$ at $M/N \geq 0.5$.
Surprisingly, this result is superior to that of the $l_1$ norm
based approach, which is known as an optimal algorithm in the
noiseless case. The gap between the two algorithms is caused by the
reconstruction error over the non-supporting elements. CS-BSD
completely removes the error from the non-supporting elements
whereas the $l_1$ norm based approach leaves a certain amount of the
reconstruction error on the non-supporting elements.

\begin{figure}[!b]
\centering
\includegraphics[width=17cm]{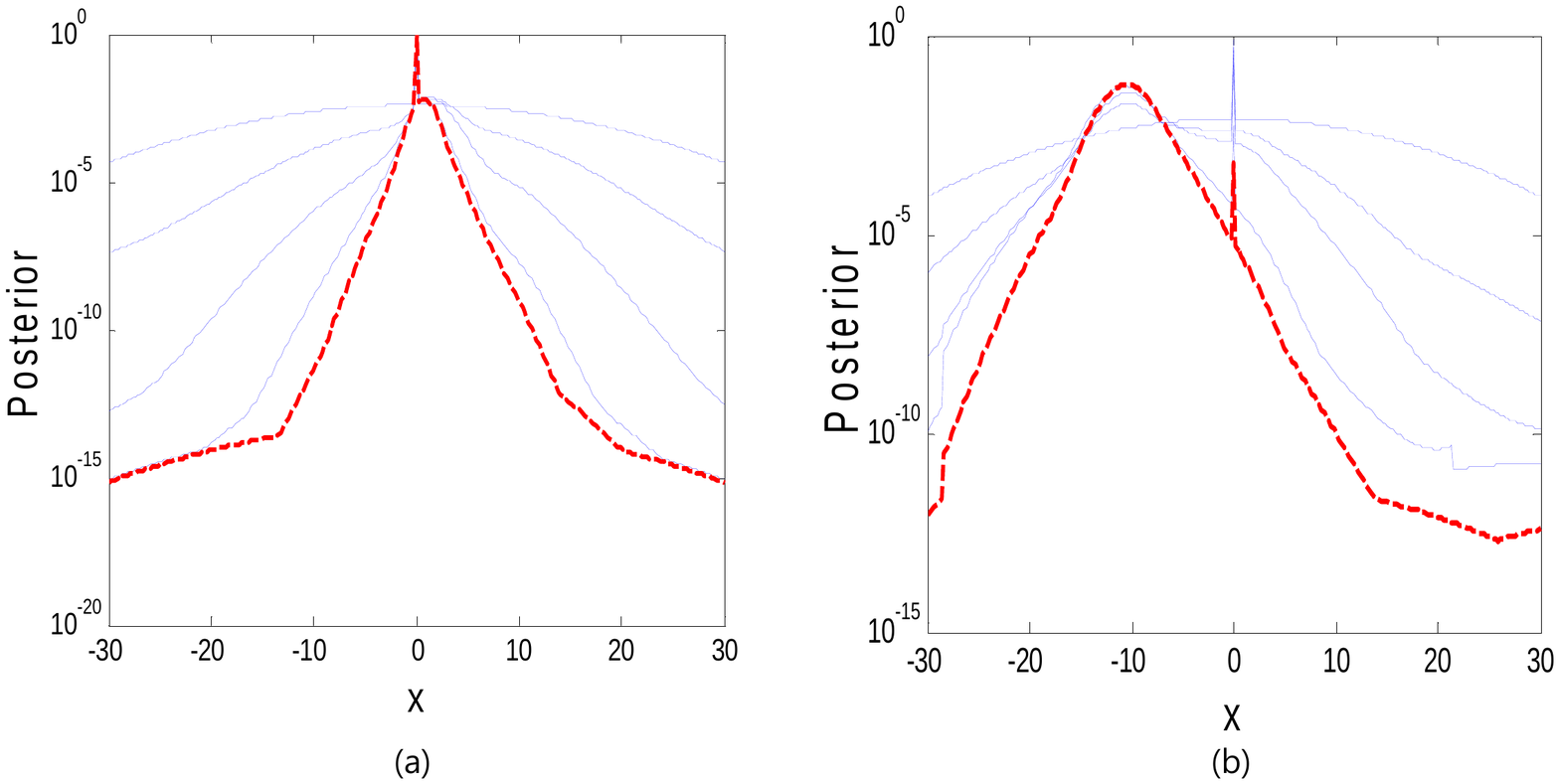}
\caption{Iterative behavior to find posterior of $x_i$ at SNR=10dB:
(a)when $ s_{0,i}=0$, (b)when $s_{0,i}=1$. The dotted-red line
indicates the posterior density after 5 iterations.}
\label{fig:Fig4-2}
\end{figure}

In the low SNR regime, it is noteworthy that CS-BSD works well
although the proposed algorithm was originally targeted at a
reasonable system having high SNR. For example, CS-BSD achieves
MSE=$10^{-2}$ at SNR=14 dB in Fig.\ref{fig:Fig5-2}, which provides 3
dB SNR gain from L1-DS; 2 dB gain from CoSaMP; 1 dB gain from CS-BP
and SuPrEM. To support this result, we present Fig.\ref{fig:Fig4-2}
which describes the iterative behavior to find the posterior of
$x_i$ given $\mathbf{z}$ at SNR=10dB. If $s_{0,i} =0$, most of the
probability mass in the posterior stays at the zero-spike as shown
in Fig.\ref{fig:Fig4-2}-(a); if $s_{0,i} = 1$, the probability mass
gradually shifts toward an estimated value as shown in
Fig.\ref{fig:Fig4-2}-(b), over the iteration. Since the SNR is low,
the probability mass spreads considerbly over the neighbored values
due to the noise effect; thus, it can lead to difficulty in
detecting the state of the signal element using the simple MAP
criterion. In CS-BSD, the use of the BHT nicely compensates for this
weakness of the MAP by scanning the probability mass over the entire
range of values.

\subsubsection{Comparison over number of iterations} 
\begin{figure}[!t]
\centering
\includegraphics[width=10cm]{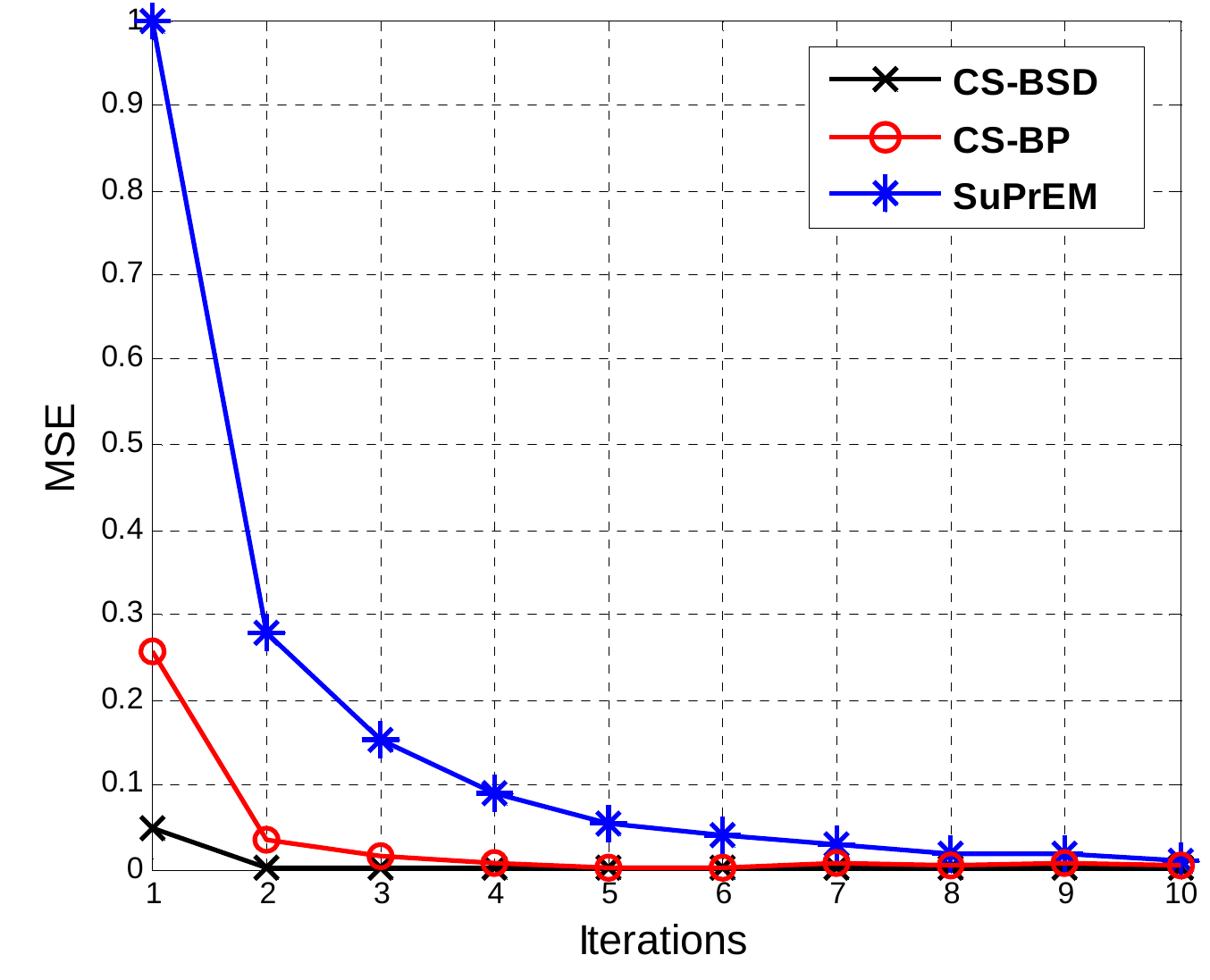}
\caption{ MSE performance of BP-based algorithms over the number of
iterations for $N=1024, M/N=0.5, q=0.1, N_d=64$, and SNR = 50
dB.}\label{fig:Fig5-5}
\end{figure}

In
Fig.\ref{fig:Fig5-5}, we examine the MSE  performance of the
BP-based algorithms, CS-BP and SuPrEM, as a function of a fixed
number of iterations where $N/M=0.5, q=0.1$, $N_d=64$, and SNR = 50
dB. In this simulation, we used the non-reweighted version of SuPrEM
since the reweighted version requires more than 10 iterations. The
figure demonstrates that CS-BSD converges faster than CS-BP and
SuPrEM. The convergence of CS-BSD is achieved within 2 to 3
iterations with CS-BP, whereas SuPrEM require more than 10
iterations.

\section{Conclusion}
The theoretical and empirical research in this paper demonstrated
that CS-BSD is a powerful algorithm for  sparse signal
reconstruction in NCS. In CS-BSD, we employed the DD estimation
structure, which consists of support detection and  signal value
estimation. In the support detection process, BP provides the signal
posterior densities, and then BHT detects the support based on the
posteriors. In the signal value estimation process, an MMSE
estimator provides the signal values using the detected support set.
These detection and estimation process are iterated until the
constraint $\mathbf{E} \| {{\bf{\Phi x}} - {\bf{z}}} \|_2 \le
\epsilon$ is met. The evaluated SER performance showed that the
support detection of CS-BSD is almost error-free beyond a certain
threshold SNR according to the undersampling ratio $M/N$. On the
basis of the SER result, we argued that CS-BSD achieves the
performance of an MMSE estimator that has the knowledge of the
support set beyond the threshold SNR. We supported the argument by
evaluating the MSE performance. The complexity of CS-BSD is $O(
N\log N +KM)$, which includes the cost of MMSE $O(KM)$, in addition
to that of BP, $O( N\log N)$. Although our algorithm incurs an
additional cost for MMSE estimation, it converges faster than other
BP-based algorithms, so the computational cost is lower in practice.

\section*{Appendix A\\Proof of Theorem 1}
\emph{Proof}: We define a random vector
$\mathbf{x}_{N_{\mathcal{C}}(j)}=[x_{N_{\mathcal{C}}(j),1},...,x_{N_{\mathcal{C}}(j),W}]$
consisting of the signal elements associated with $z_j$ and the
corresponding index set $\mathcal{W}:=\{1,...,W\}$, where
$W:=|N_{\mathcal{C}}(j)|$. With a bijective mapping function $g:
N_{\mathcal{C}}(j) \rightarrow \mathcal{W}$,  each element of
$\mathbf{x}_{N_{\mathcal{C}}(j)}$ corresponds to
\begin{eqnarray}
x_k = x_{N_{\mathcal{C}}(j),g(k)} \text{ for all}\,\,\, k \in
N_{\mathcal{C}}(j).
\end{eqnarray}
By marginalizing over $n_j$ to $f_{z_j}(z|x_i )$, we obtain
\begin{eqnarray}\label{eq:eq_App2}
{f_{{z_j}}}({z}|{x_i}) = \int\limits_{{n_j}}
{{f_{{z_j}}}({z}|{x_i},{n_j}){f_{{n_j}}}({n}|{x_i})d{n}},
\end{eqnarray}
where ${f_{{n_j}}}({n}|{x_i}) = {f_{{n_j}}}({n})$ since
 $\mathbf{n}$ is independent of $\mathbf{x}$.  By further marginalizing
over elements of $\mathbf{x}_{N_{\mathcal{C}}(j)}$, we rewrite the
expression in \eqref{eq:eq_App2} as
\begin{eqnarray}\label{eq:eq_App3}
\begin{array}{l}
{f_{{z_j}}}({z}|{x_i})
  = \int\limits_{\{x_{N_{\mathcal{C}}(j),w}\}_{w=2}^{W}} \int\limits_{n_j} f_{z_j}(z|\mathbf{x}_{N_{\mathcal{C}}(j)},n_j ) f_{n_j}(n) f_{\mathbf{x}_{N_{\mathcal{C}}(j)}|x_{N_{\mathcal{C}}(j),1}}(x_2,...,x_W|x_i)dn dx_2\cdots dx_W, \\
\end{array}
\end{eqnarray}
where we assume $x_i = x_{N_{\mathcal{C}}(j),1}$ without loss of
generality. In addition,
$f_{z_j}(z|\mathbf{x}_{N_{\mathcal{C}}(j)},n_j )=\delta(z-z_j)$
holds true since knowing $\mathbf{x}_{N_{\mathcal{C}}(j)}$ is
equivalent to knowing $ {\left( {{\bf{\Phi x}}} \right)_{row(j)} }$;
thus, there is no uncertainty in $ z_j={\left( {{\bf{\Phi x}}}
\right)_{row(j)} }+n_j$. Since the elements of $\mathbf{x}$ are
assumed be independent, we replace
$f_{\mathbf{x}_{N_{\mathcal{C}}(j)}|x_{N_{\mathcal{C}}(j),1}}(
x_2,...,x_W|x_i)$ in \eqref{eq:eq_App3} with the product of the
probability densities.
\begin{eqnarray}\label{eq:eq_App4}
{f_{{z_j}}}({z_j}|{x_i}) = \int\limits_{n_j}
{\int\limits_{\{x_{N_{\mathcal{C}}(j),w}\}_{w=2}^{W}}{\delta(z-z_j)
f_{n_j}(n) \left( {\prod\limits_{w=2}^{W}
{f_{x_{N_{\mathcal{C}}(j),w}}(x_w) } } (dx_w)\right)} dn}
\end{eqnarray}
The expression in \eqref{eq:eq_App4} can be represented by a
sequence of convolutions of probability densities, as given in
\eqref{eq:th1}.
$\,\,\,\,\,\,\,\,\,\,\,\,\,\,\,\,\,\,\,\,\,\,\,\,\,\,\,\,\,\,\,\,\,\,\,\,\,\,\,\,\,\,\,\,\,\,\,\,\,\,\,\,\,\,\,\,\,\,\,\,\,\,\,\,\,\,\,\,\,\,\,\,\,\,\,\,\,\,\,\,\,
\,\,\,\,\,\,\,\,\,\,\,\,\,\,\,\,\,\,\,\,\,\,\,\,\,\,\,\,\,\,\,\,\,\,\,\,\,\,\,\,\,\,\,\,\,\,\,\,\,\,\,\,\,\,\,\,\,\,\,\,\,\,\,\,\,\,\,\,\,\,\,\,\,\,\,\,\,\,\,\,\,
\,\,\,\,\,\,\,\,\,\,\,\,\,\,\,\,\,\,\,\,\,\,\,\,\,\,\,\,\,\,\,\,\,\,\,\,\,\,\,\,\,\,\,\,\,\,\,\,\,\,\,\,\,\,\,\,\,\,\,\,\,\,\,\,\,\,\,\,\,\,\,\,\,\,\,\,\,\,\,\,\,\blacksquare$

\section*{Acknowledgment}
This work was supported by the World-Class University Program
(R31-10026), Haek-Sim Research Program (NO. 2011-0027682), Do-Yak
Research Program (NO.2011-0016496), and Leading Foreign Research
Institute Recruitment Program (K20903001804-11E0100-00910) through
the National Research Foundation of Korea funded by the Ministry of
Education, Science, and Technology (MEST).

\begin{table}[!b]
\renewcommand{\arraystretch}{1.3}
\caption{Comparison of several recent sparse recover algorithms}
\label{table1}
 \centering

\begin{tabular}{||c||c|c|c|c||}
\hline\hline
Algorithm & Complexity for recovery        & Type of $\Phi$                      & Prior type                      & Algorithm type \\
\hline \hline
CS-BSD    & $O(N\log N+KM)$       & sparse-Bernoulli                      & spike-and-slab                   &MMSE, BP, BHT  \\
\hline
CS-BP   & $O(N\log N)$                  & sparse-Bernoulli                        & two-state Gaussian mixture     & MAP , BP \\
\hline
SuPrEM    & $O(N\log N)$               & Low-density frame                         & Jefferys', Sparsity $K$                     & MAP, BP, EM\\
\hline
BCS     & $O(NK^2)$                   & Gaussian                            & Gamma                           & MAP,  BP, EM, \\
\hline
CoSaMP   & $O(MN\log K)$             & Gaussian                             & Sparsity $K$                  & Greed pursuit\\
\hline
L1-DS  & $\Omega(N^3)$           & Gaussian                                & -                               & CVX opt. via LP\\
\hline\hline
\end{tabular}
\end{table}

\begin{algorithm}
\caption{CS-BSD}\label{alg:CS-BSD}
\begin{algorithmic}[0]
\Require Noisy measurements $\mathbf{z}$, Sensing matrix
$\mathbf{\Phi}$, Priori density $f_{x}(x)$, density of noise element
$f_{n_j}(n)$.

\Ensure Reconstructed signal $\widehat{\mathbf{x}}$, Detected
support set $\widehat{\mathbf{s}}$.
\\
\State {\bf{1)Initialization:}}

\State set $l=0$, $\epsilon$ \State set $\mathbf{b}_{j \rightarrow
i}^{l=0}=\mathbf{1}\text{ for all } (i,j) \in \mathcal{E}$

\State set $\gamma= q/(1-q) $ \While { $\mathbf{E} \| {{\bf{\Phi
}}{\bf{x}}^l - {\bf{z}}} \|_2  > \epsilon$}

    \State set $l=l+1$
    \State {\bf{2)Support Detection:}}

    \State set ${\mathbf{a}}_{i \rightarrow j}^l  = \eta [{f_{x}(x)  \times \prod\limits_{k \in N_\mathcal{V} (i)\backslash\{j\}} {{\bf{b}}_{k \rightarrow i}^{l-1}
                }} ]$, and

$\,\,\,\,\,\,\mathbf{b}_{j \rightarrow i}^l:= \delta(z - z_j)
\otimes f_{n_j}(n) \otimes \left(\bigotimes \limits_{k \in
N_{\mathcal{C}}(j)\backslash \{i\} } {\mathbf{a}_{k \rightarrow
j}^{l}} \right)$

$\,\,\,\,\text{ for all } (i,j) \in \mathcal{E}$

    \State set $f_{x_i^l}(x|\mathbf{z}) = \eta\left[ { f_{x}(x) \times
\prod\limits_{j \in N_{\mathcal{V}} (i)} {{\mathbf{b}_{j \rightarrow
i}^{l}} } } \right] \text{ for all } i \in \mathcal{V}$

    \For{$i=1$ \textbf{to} $N$}

        \If {${\frac{{\int  {\frac{{f_{x}(x |s = 0) }}{{f_{x}(x )
}}f_{x_i^l}(x |\mathbf{z} )}dx }}{{\int {\frac{{f_{x}( x |s = 1)
}}{{f_{x}( x ) }}f_{x_i^l}(x |\mathbf{z} ) }dx }}} < \gamma$} set
$\widehat{s}_i^l =1$
        \Else  {} set $\widehat{s}_i^l =0$
        \EndIf

    \EndFor
    \State set $\mathbf{\Phi}_{supp}^l(\widehat{\mathbf{s}}^l)$
    \State {\bf{3)Signal Value Estimation:}}

    \State set  $\widehat{\bf{x}}_{supp}^l  = \left(\frac{1}{{\sigma _x^2}}{\bf{I}} +
\frac{1}{{\sigma _n^2}} { {\bf{\Phi }}_{supp}^{l^{\,\,*} } {\bf{\Phi
}}_{supp}^l } \right)^{ - 1} {\bf{\Phi }}_{supp}^{l^{\,\,*} }
\frac{1}{{\sigma _n^2}}{\bf{z}}$

    \State set
    $\widehat{x}_i^l  = \left\{ \begin{array}{l}
\widehat{x}_{supp,h(i)}^l  ,\,\,\,\,{\text{if }} \widehat{s}_i =1 \\
 0,\,\,\,\,\,\,\,\,\,\,\,\,\,\,\,\,\,\,\,\,\,\,\,\,{\text{o.w. }} \\
 \end{array} \right.$ $\text{ for all } i \in \mathcal{V}$

\EndWhile
\end{algorithmic}
\end{algorithm}

\end{document}